# Effect of Deposition Pressure on the Superconductivity of Ti$_{40}$V$_{60}$ Alloy Thin Films

Shekhar Chandra Pandey[1,2], Shilpam Sharma[1], R. Venkatesh[3], L. S. Sharath Chandra[1,2] and M. K. Chattopadhyay[1,2]

[1] *Free Electron Laser Utilization Laboratory, Raja Ramanna Center for Advanced Technology, Indore, Madhya Pradesh - 452013, India*

[2] *Homi Bhabha National Institute, Training School Complex, Anushakti Nagar, Mumbai 400 094, India*

[3] *UGC-DAE Consortium for Scientific Research, Khandwa Road, Indore, Madhya Pradesh - 452009, India*

*E-mail:* shilpam@rrcat.gov.in

**Abstract.** The growth and characterization of high-quality superconducting thin films is essential for fundamental understanding and also for the use of these films in technological applications. In the present study, Ti$_{40}$V$_{60}$ alloy thin films have been deposited using DC magnetron co-sputtering of Ti and V at ambient temperatures. The effect of deposition pressure on the film morphology, superconducting and normal state properties has been studied. Measurement of electrical resistance as a function of temperature indicates that up to a certain deposition pressure, the 20 nm thick Ti$_{40}$V$_{60}$ films exhibit metallic behavior in the normal state and superconductivity at low temperatures. Beyond a threshold pressure, the films show a negative temperature coefficient of resistance with a residual resistance ratio less than one. Electrical transport measurements in the presence of magnetic field were performed to find the current-voltage (*I-V*) characteristics of the thin films. Analysis of the *I-V* curves indicates that the Ti$_{40}$V$_{60}$ alloy thin films have a large transport critical current density ($J_C$) e.g. $1.475 \times 10^{10}\ A/m^2$ in zero magnetic field and $2.657 \times 10^9\ A/m^2$ in 4 T (both at 4 K). Analysis of the field dependence of flux line pinning force density indicates a combined effect of core Δ$k$ surface and core Δ$k$ point pinning mechanisms (where $k$ is the Ginzburg-Landau parameter). Additionally, spatial variations in the superconducting critical temperature ($T_C$) across the sample contribute to $\delta T_C$ pinning. In higher magnetic fields, a contribution from $\delta l$ pinning (where $l$ is the electron mean free path) also becomes significant. The findings indicate the potential of Ti$_{40}$V$_{60}$ alloy thin film for superconducting device applications like cryogenic radiation detectors.

# Introduction

While Nb-based superconductors like NbTi and Nb$_3$Sn have long been used in high field applications such as particle accelerators, MRI, and fusion energy systems [1, 2], they face limitations in sustaining high $J_C$ in very strong magnetic fields, particularly in radiation-intensive environments like fusion reactors and space applications. Upon irradiation, Nb-based conventional superconductors degrade due to neutron-induced defects and the formation of long-lived radioactive isotopes [3-5]. This has led to the exploration of alternative superconducting materials with high $J_C$ and resilience under extreme conditions. Titanium-Vanadium (Ti$_x$V$_{100-x}$) alloys have emerged as promising candidates due to their inherent radiation resistance [6, 7] and tunable superconducting properties [8]. These alloys can withstand neutron irradiation while maintaining excellent superconducting performance, making them attractive for technological applications, especially for fusion energy research and radiation-prone environments. Several studies have explored ways to enhance $J_C$ in the Ti-V alloys, including bulk samples and multifilamentary wires [9, 10]. In arc-melted Ti$_x$V$_{100-x}$ alloys, the highest reported $J_C$ was about $10^8$ A/m² for Ti$_{70}$V$_{30}$ alloy at 4.2 K in zero field [11]. Heat treatment of Ti-rich alloys increased $J_C$ to $5 - 6 \times 10^8$ A/m² in low magnetic fields [11]. However, a sudden drop in $T_C$ for certain compositions (x = 70-82) limited their practical use [12]. Multifilamentary wires showed better performance, achieving $J_C$ up to $8.5 \times 10^8$ A/m² after annealing and cold-working [13], though still far below the widely used Nb-Ti superconductors, which exceed $5 \times 10^9$ A/m² in similar conditions [14]. To overcome these limitations, researchers have experimented with adding rare-earth elements to Ti-V alloys to improve flux pinning and increase $J_C$. Small additions of gadolinium (1 at%) raised $J_C$ to $8 \times 10^8$ A/m² [15], while 2 at% yttrium improved it to $4.8 \times 10^8$ A/m² [16]. Ti$_{40}$V$_{60}$ alloys showed higher $J_C$ values, exceeding $3 \times 10^8$ A/m² in zero field, when trace amounts of Ce, Dy, and Nd were added [12]. Further enhancements were achieved through cold-working and successive cold-working followed by annealing (SCA), pushing $J_C$ to over $4 \times 10^9$ A/m² at 4 K in zero magnetic field and improving performance even under an 8 T magnetic field [12].

As compared to the significant number of studies on the increase in the $J_C$ of the bulk Ti-V alloys, the studies on the superconducting properties of the Ti-V alloy thin films are rather scarce. Very few studies have explored the growth and characterization of Ti-V alloys in their thin film forms. Some of the reports on thin films of Ti-V alloy describe the variation in the $T_C$ for the films deposited at either cryogenic [17, 18] or high temperatures [19]. Thus, the superconducting properties of Ti-V alloy thin-films require further investigation to assess their potential for practical applications. Superconducting properties of the thin films depend on factors such as composition, deposition conditions, microstructure, and film thickness [20]. Defects, grain boundaries, and surface roughness can also significantly impact current transport and overall efficiency of the superconducting devices. The physical, morphological, structural, and optical properties of thin films, in general, are significantly affected by disorder within the system [21-23]. Disorder can be introduced by varying the deposition parameters such as the base pressure that governs the partial pressure of the residual reactive gases in the deposition system, the pressure of the sputtering gas maintained during deposition, the substrate-target distance, etc. Optimally introducing disorder can help fine-tune these properties for practical applications. In our previous work, we have studied the growth of Ti-V alloy thin films at ambient temperature with varying compositions [24] and sputtering currents [8]. We explored how these factors influence the superconductivity of Ti-V alloy thin films.

In this work, we have optimized the superconducting and normal state properties of $Ti_{40}V_{60}$ alloy thin films by tuning the sputtering gas (Ar) pressure during the deposition. The main emphasis of the present work is to control the superconducting properties of the Ti-V alloy thin films by tuning the deposition pressure. This change in deposition pressure affects the disorder in the film, which in turn influences the atomic arrangement, defect formation, and film smoothness, all of which impact the superconducting and normal state properties of the films. By systematically tuning the deposition pressure, we aim to optimize the $T_C$ and $J_C$ in the Ti-V alloy thin films in a simple and controlled manner without additional material modifications. After the deposition, the films have been characterized for structural, morphological and electrical-transport properties. Along with the superconducting parameters, we have also examined the sheet resistance ($R_S$), which is critical for superconducting device performance. Previous studies suggest that both the $T_C$ and $R_S$ can be tuned by adjusting the film thickness and substrate choice [25], though in the thicker films this effect is minimal. In some specific technological applications like superconducting nanowire single-photon detectors (SNSPDs), maintaining a consistent film thickness is essential. Our results show that by varying the deposition pressure, one can reproducibly tune the $T_C$ and $R_S$ to achieve the required values. Additionally, the transport $J_C$ of the present Ti$_{40}$V$_{60}$ thin films has been found to be high as compared to the main stream Nb-based superconducting alloys. Our study also aims to identify the dominant flux line pinning mechanisms in these thin films. Understanding these aspects is essential for assessing the potential of the Ti$_{40}$V$_{60}$ thin films in efficient and highly sensitive radiation detection applications and for a fundamental understanding of the superconductivity of Ti-V alloy thin films.

## Experimental details

The Ti-V alloy thin films were grown on SiO$_2$ (300 nm thick, amorphous) coated Si (100) substrate (10 mm × 10 mm × 0.5 mm) using a home-built DC magnetron sputtering unit. The deposition was done at ambient temperature by co-sputtering of high-purity titanium (Ti, 99.999%) and vanadium (V, 99.9%) targets from M/s Testbourne, U.K., in an ultra-high vacuum environment. The sputtering was performed using ultra-high purity argon (99.9995%) gas. In the deposition chamber, four magnetron guns (cathodes) are arranged in a confocal configuration, and the substrate holder is movable in the vertical direction to adjust the distance between substrates and targets. Before the deposition, the substrates were ultrasonically cleaned using de-ionized water, boiling acetone and ethyl alcohol and then dried at 100$^0$C. During the deposition of thin films, two out of four magnetron guns were DC powered simultaneously. To ensure the homogeneity of the film, the substrates were rotated at 25 RPM on the focal plane of the magnetron guns. The magnetron cathode was placed at a distance of about 11 cm from the substrate holder. Pre-calibrated individual deposition rates and molar volumes of Ti and V targets were used to calculate the required time for the particular thickness and alloy composition. The deposition rates of Ti and V as functions of sputtering time and sputtering current were obtained using a surface profilometer as $9.86 \times 10^{-3}$ Å$/mA.sec$ and $8.21 \times 10^{-3}$ Å$/mA.sec$ respectively. Using these deposition rates, thin films of Ti$_{40}$V$_{60}$ alloy composition with 20 nm thickness were deposited. Prior to the deposition, the sputtering system was evacuated to a base pressure of $\sim 1.8 \times 10^{-7}$ mbar. Ti$_{40}$V$_{60}$ alloy thin films were deposited at different deposition pressures by controlling the argon gas inflow using a mass flow controller from Bronkhorst, GmbH. The six thin film samples were deposited under background argon gas pressures of 1.1 (TiV-1), 1.0 (TiV-2), 0.9 (TiV-3), 0.8 (TiV-4), 0.7 (TiV-5), 0.6 (TiV-6) $\mu$bar maintained by 22, 20, 18, 16, 14, 13 sccm argon gas flow, respectively.

Phase purity of the samples was confirmed by grazing incidence X-ray diffraction (GIXRD) measurements using an X-ray diffractometer (Bruker D8 discover) with $CuK_\alpha$ radiation. The surface morphology and grain distribution of the films were studied using atomic force microscopy (AFM) in a Veeco SPM scanning probe microscope. The electrical properties of the films as a function of temperature were measured using a Cryofree Spectromag CFSM7T-1.5 magneto-optical cryostat (Oxford Instruments, UK). For the measurement of electrical resistance as a function of temperature *R(T)*, the electrical contacts were made in the four-point probe geometry, using thin copper wires and high-conductivity silver paint. Current-voltage (*I-V*) characteristics of the deposited Ti-V alloy thin films were measured in the linear bar geometry. The structures of the linear bar were made using a home-built optical photolithography set up. In this set-up a compound microscope having objective lenses of different magnification is used to focus a UV diode laser of wavelength 405 nm. The focused laser spot is rastered on the photoresist coated substrate for patterning the required structures. The *I-V* characteristics were recorded at different temperatures in magnetic fields up to 4 T. The magnetic field was applied parallel to the surface of the sample and in the direction of current flow. A Keithley 6221 current source supplied the constant current to the patterned thin films. During the measurement of *I-V* characteristics, the current was systematically varied, and the corresponding voltage was measured at various temperatures using Keithley 2182A nanovoltmeter.

## Results and discussions

Previously, we have reported the optimization of the growth and superconducting transition temperature of Ti-V alloy thin films with respect to the sputtering currents and alloy compositions [8, 24]. Results confirm the formation of the $Ti_{40}V_{60}$ composition in the body centered cubic (*bcc*) crystal structure. For the present study, we have deposited a series of six $Ti_{40}V_{60}$ alloy thin films, systematically varying the deposition pressure while keeping the sputtering current and film thickness constant (20 nm). We aimed to grow the superconducting $Ti_{40}V_{60}$ thin films with high values of sheet resistance ($R_S$), $T_C$ and $J_C$. The GIXRD measurements indicate that the $Ti_{40}V_{60}$ alloy thin films are polycrystalline in nature, regardless of the deposition pressure up to a certain limit. The diffraction patterns for all the six samples are shown in Fig. 1, where the peaks are indexed to the *bcc* crystal structure (space group, $Im\overline{3}m$).TiV-1, deposited at the highest pressure (1.1 $\mu$bar), did not show any reflection in the GIXRD measurements. This absence of peaks suggests that the film may be amorphous or very poorly ordered. High deposition pressures can create conditions that favour the formation of amorphous or non-crystalline phases [26]. Generally, the crystalline or amorphous nature of a film depends on the mobility of the ad-atoms. Low mobility can lead to a disordered or amorphous structure. Increased deposition pressure beyond a certain threshold can promote this amorphous behavior. For the other five samples, peak positions corresponding to the (110) and (211) reflections did not shift as a function of deposition pressure, indicating strong phase stability of the material and minimal stress developed during deposition.

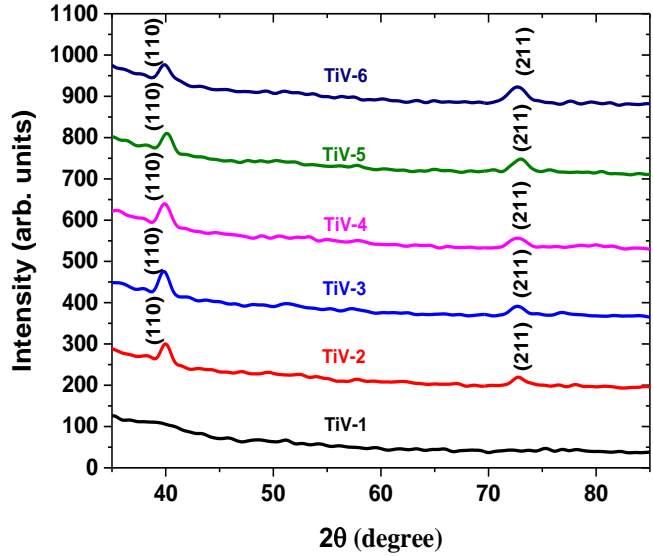

**Fig. 1**. GIXRD patterns of $Ti_{40}V_{60}$ alloy thin films deposited at 1.1 (TiV-1), 1.0 (TiV-2), 0.9 (TiV-3), 0.8 (TiV-4), 0.7 (TiV-5), 0.63 (TiV-6) $\mu$bar pressures indicating *bcc* crystal structure.

AFM imaging was done to obtain high-resolution topographical images. The scanning area was set to 0.5 $\mu$m × 0.5 $\mu$m to ensure sufficient coverage of the sample surface. The AFM micrographs reveal distinct grain boundaries and surface features, allowing for accurate measurement of grain sizes and assessment of microstructural properties. The AFM micrographs were analyzed using WSxM software [27]. The two-dimensional micrographs are shown in Fig. 2(a), 2(b) and 2(c) for the TiV-1, TiV-5 and TiV-6 samples respectively. In these panels, the grain boundaries are identified from the change of height and surface texture observed. It is observed that the films deposited at higher pressure have poor grain growth and grain connectivity. Thus, lower deposition pressures favour the conditions for grain growth, resulting in well-defined and large grains. Change in the surface morphology of the films with deposition pressure is a result of variation in the kinetic energy of the atoms impinging on the substrate. Variation in the kinetic energy is related to the mean free path of the atoms arriving at the substrate. At higher deposition pressure, the mean free path of the atoms is shorter and vice verse. Short mean free path leads to reduced kinetic energy of the ad-atoms. When the kinetic energy of the sputtered atoms is low, the surface mobility of the atoms on the substrate is not enough for the formation of large and well-connected grains. During the deposition of the films, both the power and the voltage at the targets are observed to decrease as the deposition pressure increases in the constant current mode operation of the sputtering unit. This reduction in the voltage with rising deposition pressure also leads to the low sputtering rate and kinetic energy of the sputtered atoms [28]. A reduced sputtering rate can also affect the grain and surface characteristics of the thin film. The grain distribution is crucial for the pinning of the vortices in the superconducting state and an important factor in determining the $J_C$ values.

The smoothness of the films can be quantified using its roughness. The RMS roughness of the present films were found to be in the range of 0.17 nm - 0.2 nm. The effect of deposition pressure on the roughness of the films can be related to the energy of the ad-atoms impinging on the substrate surface. The low roughness of these films can be attributed to the nano-crystalline surface morphology of the films. These low-roughness thin films are particularly beneficial for superconducting radiation detection applications due to their good optical properties and reduced electrothermal noise.

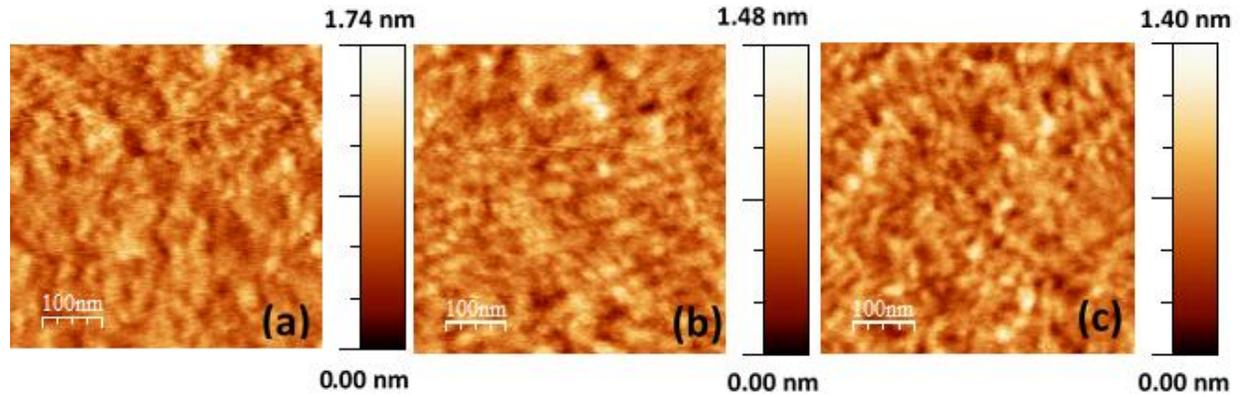

**Fig. 2**. 2D AFM micrographs of the $Ti_{40}V_{60}$ alloy thin films deposited at **(a)** 1.1 (TiV-1), **(b)** 0.7 (TiV-5), **(c)** 0.63 (TiV-6) $\mu$bar pressure showing the variation of surface morphology with deposition pressure.

The temperature dependence of the electrical resistance, $R(T)$ of the $Ti_{40}V_{60}$ alloy thin films is shown in Fig. 3(a) for the temperature range 2 to 10 K. The Inset of Fig. 3(a) displays the $R(T)$ curve for the TiV-1 and TiV-6 samples in the temperature range from 2 to 290 K. All the $R(T)$ curves presented in Fig. 3 are normalized to the resistance measured at 290 K. The nearly T-independent $R(T)$ and all the values of $R_T/R_{290K}$ close to unity observed in the present samples indicate their disordered nature. While the TiV-1 sample does not indicate any indication of superconductivity down to 2 K, all the other samples exhibit a transition to a zero-resistance superconducting state at low temperatures. It is also observed that while the temperature coefficient of resistivity (TCR) of TiV-1 is slightly negative, all the thin film samples exhibiting a normal to superconducting phase transition have a slightly positive TCR in their normal state. This point will be discussed further in the next paragraph. The $T_C$ values for the Ti$_{40}$V$_{60}$ alloy thin films, deposited under different pressures, are presented in Table 1. The *R(T)* measurements also reveal that the onset of the drop in electrical resistance in the present samples is at a much higher temperature than the global superconducting transition. The $T_C$ values in table 1 correspond to this main drop (global superconducting transition), and they vary as a function of the deposition pressure. For TiV-2, which has the lowest overall $T_C$, the above-mentioned onset is at 7.5 K [see figure 3(c)]. For TiV-6 on the other hand, where the overall $T_C$ is highest among the present films, the onset is near 12 K [see figure 3(d)]. Similar behaviour is observed for the other Ti$_{40}$V$_{60}$ thin films as well. This onset of the drop in electrical resistance at temperatures significantly higher than the overall $T_C$ might indicate the onset of weak superconducting correlations, or the existence of pre-formed Cooper pairs [29] before the actual superconducting condensation in these thin films.

**Table 1.** $T_C$, $R_S$, *RRR* and deposition pressure of the six Ti$_{40}$V$_{60}$ alloy thin film samples.

| S.No. | Sample | Deposition pressure ($\mu$bar) | $T_C$ (K) | $R_S$ ($\Omega/\square$) | *RRR* |
|---|---|---|---|---|---|
| 1. | TiV-1 | 1.1 | No $T_C$ | 86.67 | 0.96 |
| 2. | TiV-2 | 1.0 | 3.61 | 60.36 | 1.05 |
| 3. | TiV-3 | 0.9 | 4.14 | 56.80 | 1.07 |
| 4. | TiV-4 | 0.8 | 4.44 | 52.72 | 1.10 |
| 5. | TiV-5 | 0.7 | 5.72 | 41.29 | 1.12 |
| 6. | TiV-6 | 0.63 | 5.90 | 34.97 | 1.06 |

In this work, we have observed that reducing the deposition pressure leads to an increase in the $T_C$ of these films. In general, the disorder effect occurs in granular films, where electrical conduction between adjacent grains occurs either by Josephson tunnelling through weak links or via quasi-particle tunneling [30]. As a result, the density of grain boundaries is a crucial factor for electronic conduction in the granular systems. The observed rise in the $T_C$ of the Ti$_{40}$V$_{60}$ alloy thin films with decreasing deposition pressure is thus attributed to the improved grain connectivity. Enhanced grain growth correlates with increased sputtering rate as the deposition pressure decreases. Furthermore, at pressures above 1 µbar, the Ti$_{40}$V$_{60}$ alloy thin films show a negative TCR. The residual resistance ratio $\left(RRR = R_{290K}/R_{1.1\,T_C}\right)$ of all the films are also presented in table 1. Since TiV-1 does not exhibit superconductivity, the electrical resistance at 2

K (lowest achievable temperature in the set-up) has been used to find the *RRR* value for this sample. The film deposited at the highest deposition pressure (TiV-1) has *RRR* value below one, while the other thin films have *RRR* values slightly more than one. This indicates that the thin film deposited at the highest deposition pressure behaves like a typical dirty metal, where defect and grain boundary scattering play significant roles [31]. The *RRR* values correlate well with the $T_C$ values. The higher $T_C$ is associated with the more metallic behaviour. However, the film deposited at the lowest deposition pressure, TiV-6, exhibits a dip in the *RRR* as compared to TiV-5. At very low pressure, the growth rate may decrease significantly leading to an increase in defects or changes in the deposition environment that can negatively impact the conductivity. For the disordered superconducting thin films, the sheet resistance $R_S$ of the film is very crucial for their application in superconducting radiation detectors [32]. Materials with high sheet resistance and high normal state resistivity are suitable for these detectors. As shown in table 1, the sheet resistance of the films increases with higher deposition pressure. But, as we mentioned earlier, the $T_C$ of the films decreases as the deposition pressure increases. Therefore, optimizing the Ti-V alloy thin films to achieve reasonably high values of both $T_C$ and $R_S$ is essential. The variation of the $T_C$ and $R_S$ of the Ti$_{40}$V$_{60}$ alloy thin films with deposition pressure is shown in Fig. 3(b). The figure shows that all the films maintain high $R_S$ values with adequately high $T_C$. However, a deposition pressure of $8 - 9 \times 10^{-4}$ mbar seems to produce the best optimized Ti-V thin film superconductors for the present purpose.

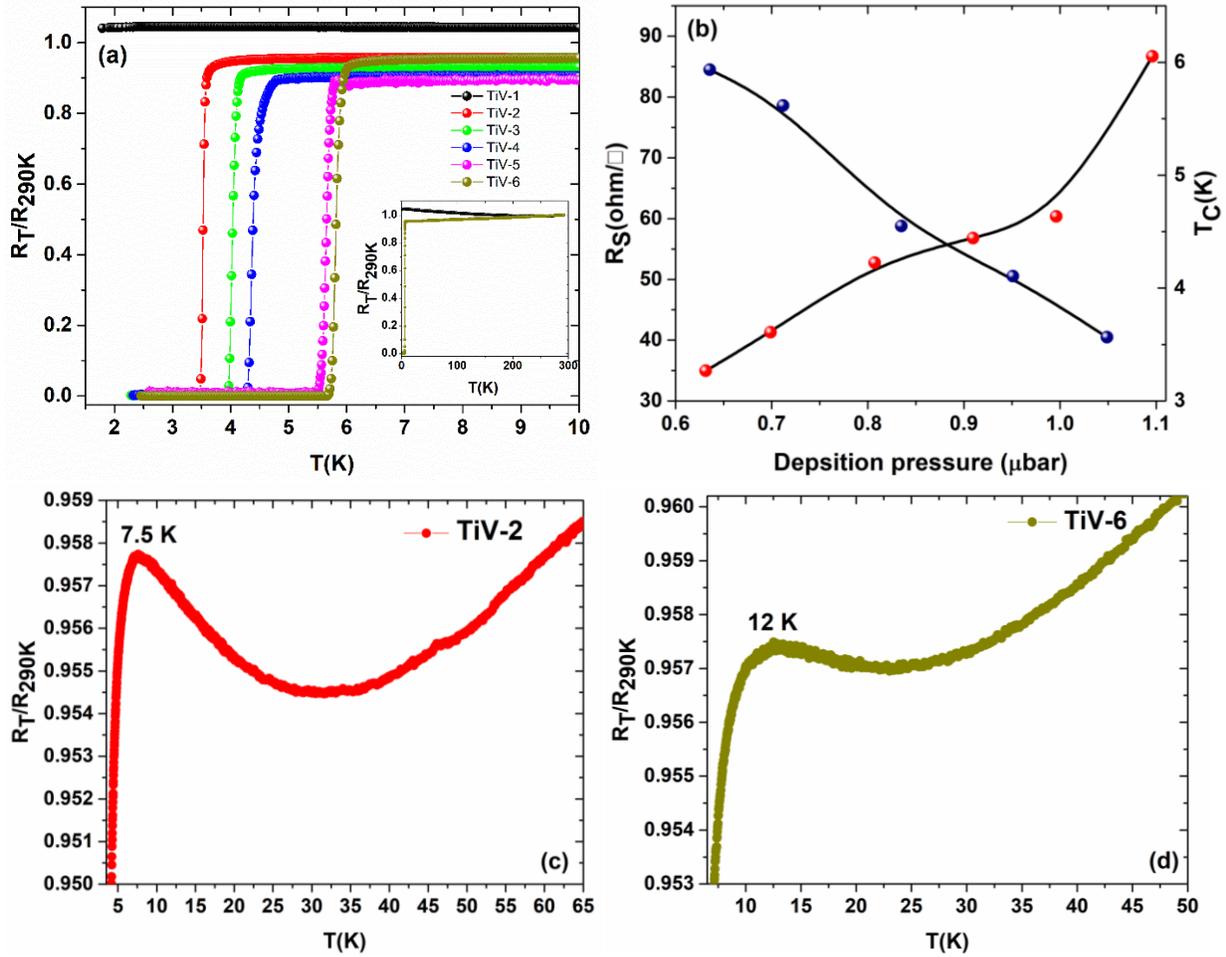

**Fig. 3 (a)** Normalized *R(T)* for six Ti-V thin films in the temperature range 2 to 10 K. The inset shows the normalized *R(T)* behavior of the films with highest and lowest deposition pressures in the temperature range 2 to 290 K. **(b)** Variation of $T_C$ and $R_S$ with deposition pressure. **(c)** Onset of the drop in *R(T)* for TiV-2 at 7.5 K. **(d)** Onset of the drop in *R(T)* for TiV-6 at 12 K.

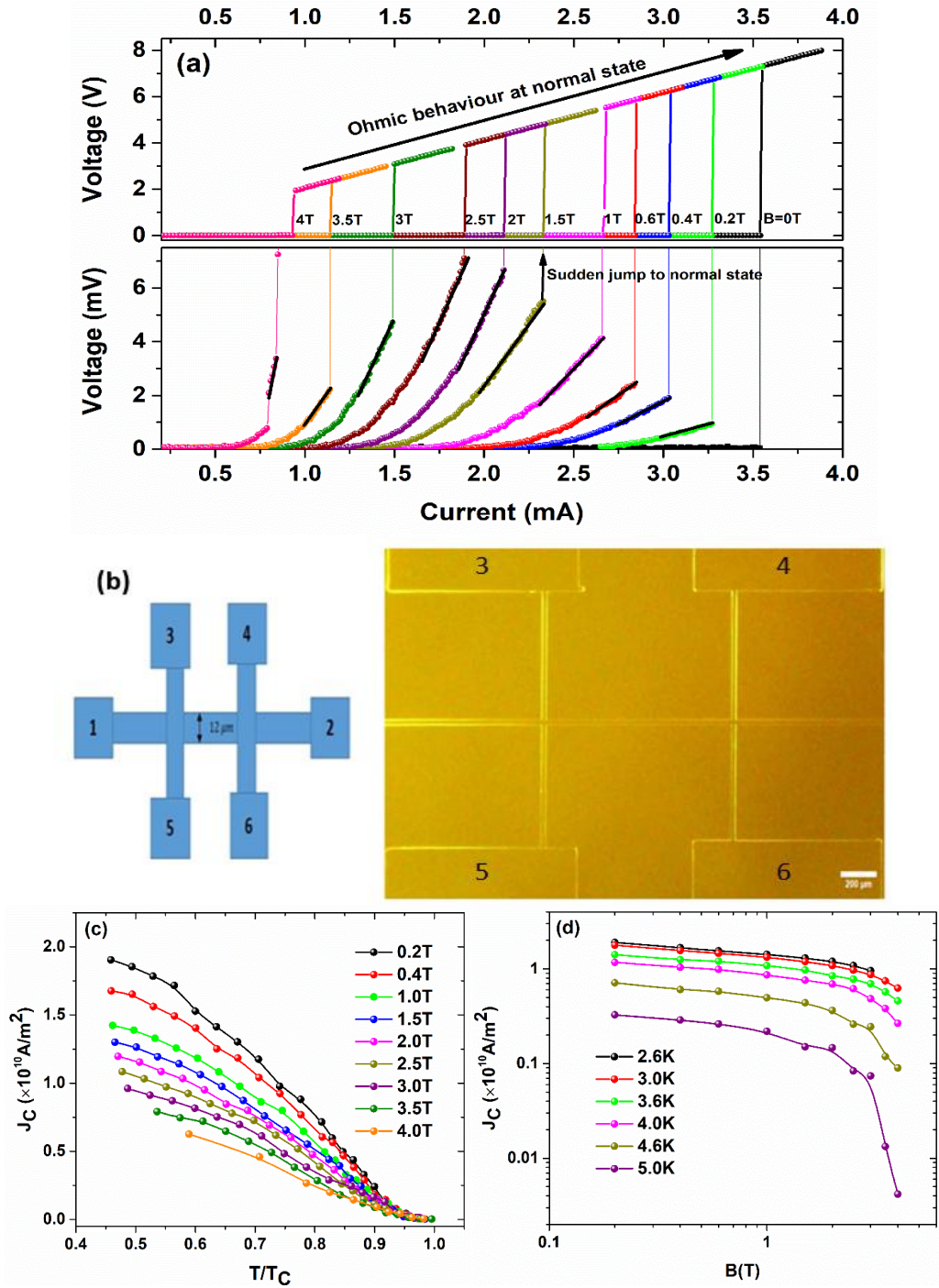

**Fig. 4 (a)** Isothermal *I-V* characteristics of TiV-6 in different magnetic fields at 4 K showing dissipationless, flux-motion and normal conducting regimes. **(b)** Hall bar geometry 12 *μ*m line width, where current leads are connected to contact pads 1 and 2, and the voltage is measured using contact pads 5 and 6. The left panel shows the schematic, while the right panel presents an image of the fabricated structure. **(c)** Variation of critical current density with reduced temperature in different magnetic fields. **(d)** Variation of critical current density with magnetic field at different temperatures in the log-log scale.

The $J_C$ measurement of the TiV-6 sample, performed in the linear bar geometry with 12 $\mu m$ line width, was done by recording the *I-V* characteristics (see Fig. 4(a)). Fig. 4(b) shows a schematic of this geometry (left side) and a photograph of the actual structure (right side) developed using photolithography. The current leads are connected to the contact pads 1 and 2, while the voltage is measured using contact pads 5 and 6. Fig 4(a) illustrates the *I-V* characteristic of the Ti-V alloy thin film TiV-6 at 4 K. The measurement was done with increasing current in different magnetic fields (parallel to the sample surface) up to 4 T. Magnetic flux lines penetrate the type-II superconductors and get pinned at certain locations at the defect sites [33]. When the applied current is low, all the vortices remain pinned, resulting in zero voltage drop across the sample. At a large enough current, the vortices begin to move because of the Lorentz force and a voltage appears across the sample. As seen in figure 4(a), the *I-V* characteristics rise very gradually from zero voltage, show a rounding behavior and then increase rapidly and linearly ($V \propto I$) with increasing *I*. With further increase of *I*, the voltage rises discontinuously to a higher value and then slowly rises linearly again with increasing *I*. The linear *I-V* regime below the discontinuous jump is used to find the $J_C$ of the present samples. We fitted a straight line to this linear *I-V* regime and extrapolated it to the current axis ($V = 0$). The corresponding current (intersection) is taken as the critical current $I_C$ [34], and the $J_C$ is obtained therefrom using the sample geometry. The solid black lines in the *I-V* curves of figure 4(a) represent the region of linear fitting. The critical current density for the Ti-V alloy thin film in linear bar geometry, with a thickness of 20 nm and a width of 12 $\mu$m, was found to be $1.475 \times 10^{10}$ and $2.657 \times 10^9$ $A/m^2$ for 0 and 4 T applied magnetic fields respectively, at 4 K. These values are significantly higher than the bulk Ti-V alloys [11] and comparable to the materials reported for SNSPD device fabrication [35]. The temperature dependence of the transport $J_C$ in various magnetic fields up to 4 T is shown in Fig. 4(c). The $J_C$ is plotted as a function of the reduced temperature $\left(t = T/T_C\right)$. The black solid lines represent the dependence of the transport $J_C$ on *t*, according to equation,

$$J_C = J_{C0}(1-t^2)^p(1+t^2)^q \qquad (1)$$

where, $J_{C0}$ is the critical current density at $T = 0$ and *p* and *q* are the exponents. Similar dependence of transport $J_C$ on *t* has been reported for type-II superconductors and linked to the pinning mechanisms [36, 37]. Pinning mechanism in the present sample will be discussed later on in this report. Fig. 4(d) shows the variation of $J_C$ with magnetic field at different temperatures ($J_C(B,T)$), plotted on logarithmic scale. The $J_C$ (*B,T*) curves indicate that the $Ti_{40}V_{60}$ alloy thin film can carry a large dissipationless current in zero as well as high fields. The high values of $J_C$ and tunability in terms of the $T_C$ and $R_S$ values makes the Ti-V alloy thin films promising materials for the efficient and sensitive superconducting radiation detectors like SNSPDs and lumped element kinetic inductance detectors (LEKIDs). Initially, the variation of $J_C$ exhibits a plateau (nearly constant behavior) in lower to moderate magnetic fields, then gradually decreases and finally exhibits a sudden fall as the magnetic field increases beyond a threshold value. This type of behavior was well reported in the literature and attributed to a transition from single vortex pinning to flux bundle pinning regimes [38]. The plateau region in low magnetic field occurs because the superconductor can effectively expel the magnetic field, allowing it to maintain superconductivity with minimal flux line penetration and strong pinning. However, as the magnetic field increases, a greater number of flux lines penetrate the superconductor, leading to a significant reduction in $J_C$. The field strength where the curve starts to drop, decreases as the temperature increases. This indicates that at higher temperature, the increased thermal energy provides sufficient excitation for depinning the flux lines quite easily.

The flux-line pinning mechanism in superconductors is significantly influenced by temperature, magnetic field and the microstructure and it has been observed that vortex motion strongly affects the detection mechanism of SNSPDs [39]. In Fig. 5(a), we present the reduced pinning force $\left(\frac{F_P}{F_{Pmax}}\right)$ as a function of reduced field $\left(\frac{H}{H_{irr}}\right)$ at 4 K (solid black symbol). The pinning force $F_P$, can be written as $F_P = J_C \times \mu_0 H$, and $F_{Pmax}$ is the maximum pinning force at a particular temperature. $H$ is the applied magnetic field and $H_{irr}$ is the irreversibility field. The red dashed lines indicate the fitting of the experimental data using Dew-Hughes formalism, taking into account the mixed effect of core $\Delta k$ surface pinning and core $\Delta k$ point pinning [33]. Here $k$ is the Ginzberg-Landau (GL) parameter. Equation 2 incorporates the contribution of these two factors for fitting the experimental data, and indicates that the Ti$_{40}$V$_{60}$ alloy thin film possesses multiple flux-line pinning mechanisms.

$$\frac{F_P}{F_{Pmax}} = A \left(\frac{H}{H_{irr}}\right)^{3/2} \left(1 - \frac{H}{H_{irr}}\right) + B \left(\frac{H}{H_{irr}}\right)^2 \left(1 - \frac{H}{H_{irr}}\right) \qquad (2)$$

Here, $A$ and $B$ are the constant, which scales the contribution of two different pinning mechanisms. In the Dew-Hughes formalism, $\Delta k$ pinning results from the spatial variation in $k$, and occurs when there are inhomogeneities in the superconducting properties. The interaction between the flux lines and the pinning centers arises due to differences in the superconducting properties between the pinning centers and the bulk of the superconductor. Small variations in $k$, caused by compositional fluctuations, changes in normal-state resistivity, and uneven dislocation distribution, contribute to $\Delta k$ pinning [33]. If the size or spacing of the pinning centers is smaller than the superconducting penetration depth, the magnetic induction ($B$) averages out within the superconductor. This results in a different free energy for the normal cores as compared to the rest of the superconducting matrix. This is known as the core interaction. The nature of $\Delta k$ pinning, whether point or surface, depends on the dimensions of the pinning centers. If their size in all directions is smaller than the inter-flux-line spacing, they are called point pins. If their size in two directions exceeds the inter-flux-line spacing, they are referred to as surface pins [33].

The maximum $T_C$ reported in the bulk $Ti_{40}V_{60}$ alloys is 7.4 K [11]. However, as per band structure calculations, the maximum $T_C$ may be much higher [40]. Previous papers from our laboratory have argued that the lower experimental $T_C$ found in these alloys is due to the presence of spin fluctuations [40]. However, Hermann J. Spitzer reported that the highest $T_C$ observed in $Ti_{13}V_{87}$ alloy thin films was 12.8 K [19], achieved when the films were deposited and annealed at high temperatures (700°C). As previously mentioned, the $Ti_{40}V_{60}$ alloy thin films exhibit a decrease in resistance at higher temperatures, with the global $T_C$ being found at lower temperatures. For the TiV- 6 sample, the $T_C$ onset occurs at 12 K, but the global $T_C$ is observed at 5.9 K. This initial $T_C$ onset may be attributed to the disorder induced destruction of the coherence of the spin fluctuations and the presence of pre-formed Cooper pairs. As the temperature decreases, more Cooper pairs are formed and finally their condensation leads to a zero-resistance state. The resistance drop at temperatures higher than global $T_C$ is also an indication of the variation of resistance and $T_C$ in the sample at different microscopic regions. We argue that the boundary between two such superconducting regions causes the core $\Delta k$ surface pinning. As the present Ti-V alloy thin film samples are polycrystalline in nature, there is a possibility for some nano size grains or small crystallites to become superconducting at different temperatures. Such regions contribute to the core $\Delta k$ point pinning.

Another classification of vortex pinning is based on the variation of $J_C$ with reduced temperature (*t*), as described in Equation 1. This variation is linked to either the spatial variation in the charge carrier mean free path (*l*) or the critical temperature ($T_C$). These two variations correspond to distinct pinning mechanisms known as $\delta l$ and $\delta T_C$ respectively [38]. The $\delta l$ pinning occurs due to spatial variations in the charge carrier mean free path around normal particles or lattice defects, while $\delta T_C$ pinning originates from the spatial variation of the GL parameter associated with fluctuations in $T_C$. The impact of these variations on the GL parameter and the GL free energy function are well documented in the literature [36]. Fig. 5(b) shows the normalized critical current density $\left(\frac{J_C}{J_{C0}}\right)$ as a function of reduced temperature. The critical current density is normalized to the critical current density at zero temperature ($J_{C0}$). The $J_{C0}$ is determined by fitting the data with the help of equation 1, using $J_{C0}$, $p$ and $q$ as the fitting parameters. The black and red solid lines represents the variation of $\left(\frac{J_C}{J_{C0}}\right)$ according to the theoretical predictions for $\delta l$ and $\delta T_C$ pinning mechanisms, as formulated in the equation 3 and 4 respectively [38].

$$J_C = J_{C0}(1 - t^2)^{5/2}(1 + t^2)^{-1/2} \qquad (3)$$

$$J_C = J_{C0}(1-t^2)^{7/6}(1+t^2)^{5/6} \qquad (4)$$

Figure 5(b) is the plot of the normalized $J_C$ as a function of reduced temperature in different magnetic fields. $\delta T_C$ pinning is the dominant pinning mechanism in the 20 nm thick $Ti_{40}V_{60}$ alloy thin film. Interestingly, in low magnetic fields, the flux line pinning is mostly governed by the $\delta T_C$ pinning while a small contribution from $\delta l$ pinning arises for $T/T_C > 0.85$, where the pinning strength is very low. This region mostly correlates with the high magnetic field data.

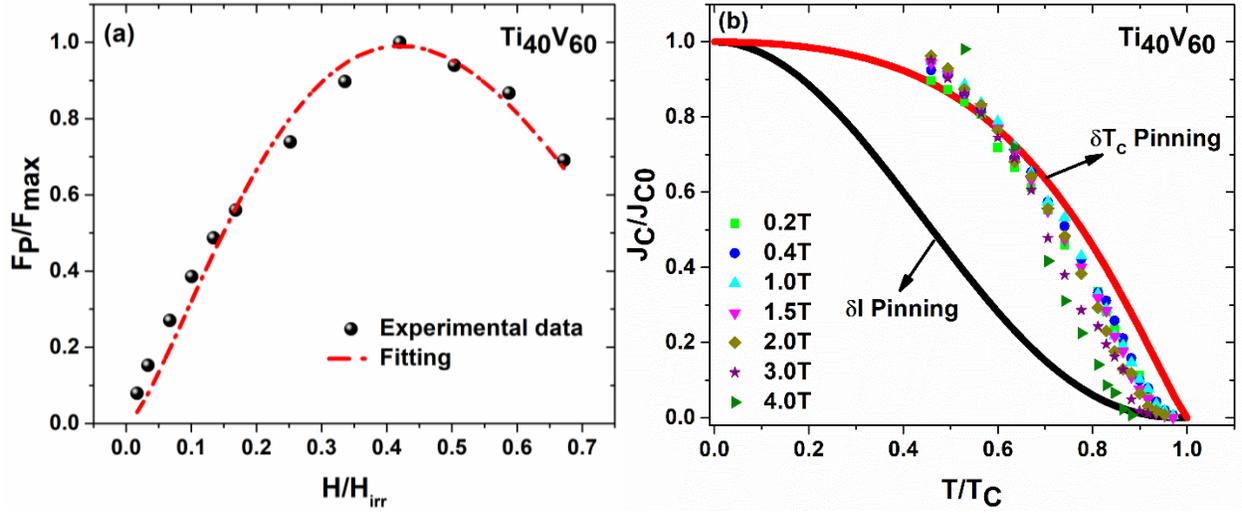

**Fig. 5 (a)** Normalized pinning force as a function of reduced magnetic field at 4 K temperature. The solid dots represent the experimental data and the dashed red line indicates the fitting of the experimental data using Dew-Hughes formalism. **(b)** Normalized critical current density as a function of reduced temperature in various magnetic fields. The solid black and red lines indicate the standard curve for $\delta l$ and $\delta T_C$ pinning respectively. $\delta T_C$ pinning is observed to be the dominant pinning mechanism in the sample.

The analysis of the pinning mechanism indicates that in the 20 nm $Ti_{40}V_{60}$ alloy thin film, pinning arises from spatial variations in superconducting properties associated with variation in $T_C$, which is influenced by surface and point defects. For effective vortex pinning in thin films, defects comparable in size to the coherence length should be considered optimal pinning sites. During the thin film deposition process, small islands initially form, which then coalesce to create a smoother film. However, these islands introduce disorder among the grains and alter the surface characteristics, leading to strong pinning centers within the film [34]. The effects of this surface modulation depend on the film's thickness. As the thickness decreases, the coalescence of the islands becomes less uniform, resulting in small amorphous regions at some grain boundaries. This kind of behavior is expected to be prominent in our samples, as the thickness is low (20 nm). These amorphous regions, with varying sizes and densities, contribute to fluctuations in the $T_C$ and lead to $\delta T_C$ pinning as well as spatial variations in the superconducting properties which contributes to the $\Delta k$ surface and point pinning. Additionally, polycrystalline films consist of grains separated by grain boundaries, with variations in crystallite size, orientation, and defect density. These features introduce spatial variations in the local electronic properties, including the mean free path (*l*) of the electrons and contributes to the *δl* pinning.

As discussed earlier, the *R(T)* measurement shows a resistance drop at higher temperatures, with a global $T_C$ observed at slightly lower temperatures. The influence of pinning in different magnetic fields can be explained as follows: In low magnetic fields, the $T_C$ contributions come from both the low $T_C$ and high $T_C$ regions, which contributes to $\delta T_C$ pinning. However, as the magnetic field increases, the contribution from the low $T_C$ regions are suppressed, making the contribution of *δl* pinning observable.

## Summary and Conclusion

We have deposited the $Ti_{40}V_{60}$ alloy thin films in $SiO_2$ coated Si substrate using DC magnetron sputtering. We have deposited the films under six different deposition pressures to optimize the $T_C$ of the films and to study how deposition pressure affects the superconducting and normal state properties of the film. The present study shows that as the deposition pressure increases, the $T_C$ of the Ti-V alloy thin films decreases and the sheet resistance increases. The *RRR* values of these films indicate that, above a certain threshold deposition pressure, the film develops a negative temperature of coefficient resistance and *RRR* value less than 1. By controlling the deposition pressure, the electronic conduction changes from good metallic state to bad metallic state (*RRR* changes from greater than 1 to less than 1), which affects the sheet resistance. Additionally, XRD results reveal that the film deposited under low Ar-pressure is polycrystalline and the film grown under highest deposition pressure is amorphous in nature. $8 - 9 \times 10^{-4}$ mbar deposition pressure seems to produce the best optimized $Ti_{40}V_{60}$ thin films with high values of $T_C$ and critical current density. *I-V* measurements reveal the flux line pinning mechanism in the 20 nm thick $Ti_{40}V_{60}$ alloy thin films and the critical current density is calculated by linear fitting of the *I-V* curve prior to the jump in normal conducting state. We have found high $J_C$ values in zero as well as high magnetic fields. Temperature and magnetic field dependence of $J_C$ was also studied and the analysis indicates that the spatial variation of $T_C$ in the films is the primary reason for the flux line pinning in the sample. The calculated values of the $J_C$ is $1.475 \times 10^{10}$ and $2.657 \times 10^9 \, A/m^2$ in 0 and 4 T applied magnetic fields respectively, both at 4K.

In conclusion, sputter deposited Ti$_{40}$V$_{60}$ thin films with varying deposition pressure are advantageous for superconducting device fabrication, as they allow precise adjustments to the $T_C$ and $R_S$ of the superconducting material. Additionally, the ability to fine tune $T_C$, along with high values of critical current density and sheet resistance, are likely to boost the utility of these alloys in technological applications. Tuning of the deposition pressure influences the defect levels and surface properties. Improving surface characteristics and reducing defects will help to reduce noise in the superconducting devices such as SNSPDs and LEKIDs. The capacity to control these factors through deposition techniques facilitates scalable production without compromising the quality and performance, which is vital for commercial applications of superconductors.

## Acknowledgements

We thank Dr. S. K. Rai, RRCAT Indore for his help in XRD measurements and Mr. Mohan K. Gangrade, UGC - DAE Consortium for Scientific Research for the help in AFM measurements.